%% file: main.tex
\documentclass[a4paper, amsfonts, amssymb, amsmath, reprint, showkeys, nofootinbib, twoside]{revtex4-1}
\usepackage[english]{babel}
\usepackage[utf8]{inputenc}
\usepackage{printlen}
\usepackage[detect-all]{siunitx}
\usepackage{comment}
\usepackage[normalem]{ulem} 
\usepackage[colorinlistoftodos, color=green!40, prependcaption]{todonotes}
\input{preamble}
\usepackage[pdftex, pdftitle={Article}, pdfauthor={Author}]{hyperref} 

\usepackage{soul,xcolor}

\begin{document}
\setstcolor{red}
\title{Optimization of laser stabilization via self-injection locking to WGM microresonator}

\author{Ramzil R. Galiev\textsuperscript{1,2}}
\email{ramzil.galiev@gmail.com}
\author{Nikita M. Kondratiev\textsuperscript{1}}
 \email{noxobar@mail.ru}
\author{Valery E. Lobanov\textsuperscript{1}}%
\author{Andrey B. Matsko\textsuperscript{3}}
\author{Igor A. Bilenko\textsuperscript{1,2}}
\affiliation{\textsuperscript{1}Russian Quantum Center, 143025 Skolkovo, Russia
}
\affiliation{\textsuperscript{2}Faculty of Physics, Lomonosov Moscow State University, 119991 Moscow, Russia}
\affiliation{\textsuperscript{3}Jet Propulsion Laboratory, California Institute of Technology, Pasadena, CA 91109-8099 USA}
\date{\today} 

\begin{abstract}
Self-injection locking is a dynamic phenomenon representing stabilization of the emission frequency of an oscillator with a passive cavity enabling frequency filtered coherent feedback to the oscillator cavity. For instance, self-injection locking of a semiconductor laser to a high-quality-factor (high-Q) whispering gallery mode (WGM) microresonator can result in multiple orders of magnitude reduction of the laser linewidth. The phenomenon was broadly studied in experiments, but its detailed theoretical model allowing improving the stabilization performance does not exist. In this paper we develop such a theory. We introduce five parameters identifying efficiency of the self-injection locking in an experiment, comprising back-scattering efficiency, phase delay between the laser and the high-Q cavities, frequency detuning between the laser and the high-Q cavities, the pump coupling efficiency, the optical path length between the laser and the microresonator. Our calculations show that the laser linewidth can be improved by two orders of magnitude compared with the case of not optimal self-injection locking. We present recommendations  on the experimental realization of the optimal self-injection locking regime. The theoretical model provides deeper understanding of the self-injection locking and benefits multiple practical applications of self-injection locked oscillators.
\end{abstract}


\maketitle

\section{Introduction}
Self-injection locking phenomenon is one of profound effects observed in oscillatory circuits. For many years this effect has been used in Radio-Physics, Radio-Engineering and Microwave Electronics with the goal to improve the spectral purity of the devices \cite{Ohta1,Ohta2,Ota1,Chang1,Chang2,magnetron1,magnetron2,gyrotron1,gyrotron2,6999932,7119883}. It also has been widely applied for stabilization of laser sources enabling various practical applications, including high resolution spectroscopy and high-precision metrology. Self-injection locking of a chip-scale semiconductor laser to an optical microcavity results in the sub-kHz laser linewidth, orders of magnitude smaller than the original linewidth of the semiconductor lasers \cite{Liang:10,Liang2015}. In this work we develop a theory that, on one hand, enables deeper understanding of the salient physical features of the self-injection locking, and, on the other hand, allowing improving the experimental results. The theory also elucidates the fundamental limitations of the linewidth of the self-injection locked oscillators.

Self-injection locking of oscillators was extensively studied for last thirty years. It was shown initially that adding an extra partially transparent mirror at the output of a Fabry-Perot laser can lead to the  noise reduction of the laser \cite{osti_6336783,Lang1980,Belenov_1983,Patzak1983,Agrawal1984,Tkach1986}. However, this stabilization scheme has significant limitations due to the dynamic instability arising with the strong enough optical feedback. The relative power feedback at the level of $10^{-4}$ is able to destabilize the system. 

The instability can be reduced if the feedback is frequency selective. Locking a laser generation line to a high-quality-factor (high-Q) mode of an external resonator provides fast frequency-selective optical feedback, which leads to improved stabilization of the laser frequency \cite{Dahmani:87,Hollberg:1987,Li:1989,hemmerich90oc,Hemmerich:94}. This configuration is dynamically stable and can produce coherent light even when the relative power feedback exceeds tens of percent. It was initially demonstrated with vacuum ring cavities \cite{Dahmani:87}. More recently it was studied with monolithic cavities such as the total internal reflection resonators (TIRRs) \cite{Hemmerich:94}. It was shown that the locking results in reduction of the phase and amplitude noise \cite{Dahmani:87,hjelme91jqe}, allows frequency tuning the laser emission \cite{Hemmerich:94}, and also facilitates efficient frequency doubling \cite{Hemmerich:94}. The laser linewidth can be improved by six orders of magnitude if a high quality factor microresonator is involved \cite{Liang2015}. 

A theory of the self-injection locking phenomenon was developed for larger optical cavities nearly thirty years ago \cite{Li:1989,hjelme91jqe}. The analysis indicated that to achieve the best performance one needs to have high Q-factor of the optical modes, low modal density, and highly stable optical path. Unfortunately, the stabilization technique using the large optical cavities was not frequently utilized because of the sensitivity of the cavities to the conditions of the experiments.

Whispering gallery mode microresonators (WGMRs) \cite{BRAGINSKY1989393,PhysRevA.70.051804,1588878,Savchenkov:07,doi:10.1002/lpor.201000025,Lin:14,Henriet:15,strekalov:2016,GRUDININ200633,Lecaplain2016MidinfraredUR,Shitikov:18}, combining high quality factor in a wide spectral range with small size and low environmental sensitivity, have proven to be suitable elements for implementing self-injection locking approach.  Recent studies have demonstrated the possibility of using high-Q optical WGMRs for passive stabilization of single-frequency \cite{VASSILIEV1998305,Vassiliev2003,Liang:10,Liang2015,Dale:16,Savchenkov:19a} or even multifrequency \cite{Pavlov2018,Donvalkar_2018,Galiev:18,Pavlov_18np,Savchenkov:19} semiconductor lasers to sub-kilohertz linewidths. Some of the lasers became commercial products \cite{OEwaves}. Very recent studies have shown a possibility of assembling the lasers on photonic integrated circuits, where the WGMRs were replaced with the high-Q microrings \cite{Stern2018,Li:18,Gaeta2019,Raja2019,Raja:20,Kovach:20}.

Despite the excellent experimental results, a systematic analysis of the optimal parameters of the self-injection locking using WGMRs has not been carried out yet. The preliminary studies \cite{oraevsky01qe,Kondratiev:17} considered oversimplified models that do not take into account all the parameters of the complex system. In this work we introduce five main parameters: a) the coupling strength of the forward and backward waves defined by the backscattering in the resonator and the associated feedback efficiency; b) the locking phase determined by the optical path between the laser and the microresonator and the frequency of the microresonator locking mode; c) the optical path between the laser and the microresonator itself; d) the laser cavity (LC)-microresonator frequency detuning, defining the working point of the system, and e) the pump coupling efficiency to the resonator mode defined by the geometrical mode matching (note, that the last four parameters can vary and can be defined in an experiment). We study in detail self-injection locking for a wide range of these parameters and show that there exists a global optimum for four of them. We show, for instance, that the increase of the backscattering coefficient above some optimal value does not provide better stabilization. We also offer recommendations on reaching the optimal frequency detuning and phase delay for the laser cavity and WGMR. Our analysis shows that the optimization allows improving the laser linewidth by orders of magnitude if compared with the not optimally selected parameters.

The paper is organized as follows. The theoretical model and basic equations are introduced in Section II. The results of the analytical and numerical optimization of the system are presented in Section III. Methods of the experimental realization of the optimal self-injection locking regime and limitations of the developed model are discussed in Section IV.
Section V concludes the paper.

\section{Self-injection locking model and basic parameters}

\begin{figure}[ht]
\centering
\includegraphics[width=.99\linewidth]{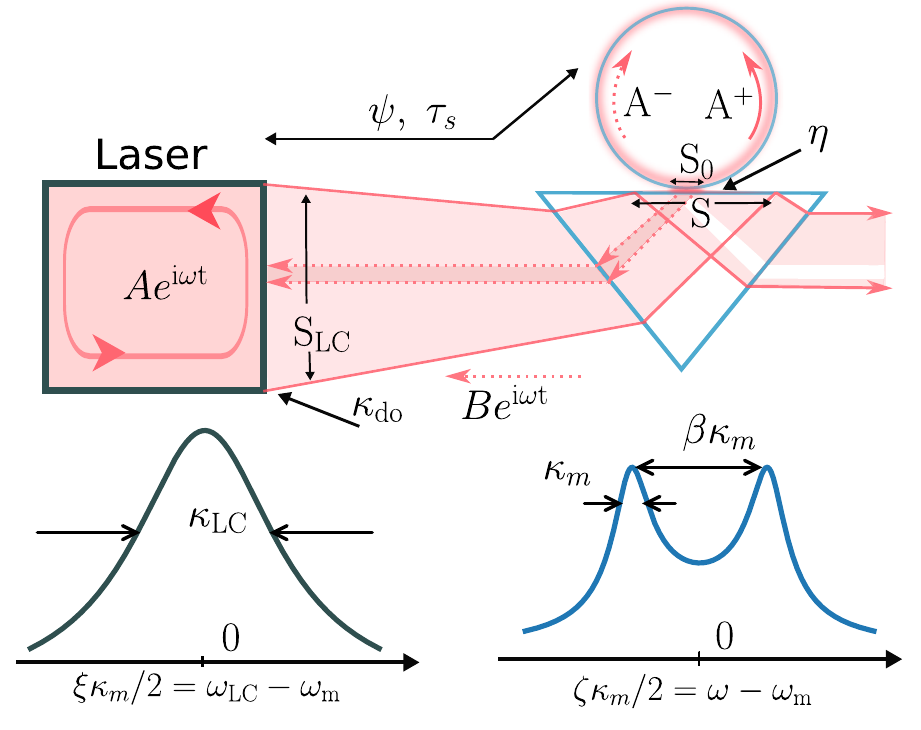}
\caption{
 Top: Scheme of a self-injection locking to a WGMR mode by means of a coupling prism. Bottom: Resonance curves of the laser (left) and the WGMR with normal mode splitting (right). $A$ -- laser generation field amplitude, $\omega_{\rm LC}$ and $\kappa_{\rm LC}$ -- the laser cavity mode frequency and linewidth, $\kappa_{\rm do}$ is laser output mirror coupling rate, $B$ - back-reflected wave, $S_{\rm LC}$ and S -- laser beam cross section area at the laser aperture and on the prism surface, $\tau_s$ - round-trip time of feedback, $\psi$ -- locking phase, $A^{+}$ and $A^{-}$ -- amplitudes of the forward and backward waves inside microresonator, $\eta$ -- microresonator coupling coefficient, $\omega_m$ and $\kappa_m$ -- the microresonator mode frequency and linewidth, $\omega$ -- generation frequency.} 
\label{fig:scheme}
\end{figure}

The schematic description of the self-injection locking effect is presented in Fig.~\ref{fig:scheme}, where a refocused laser beam is resonantly coupled to a high-Q WGM resonator. Due to the Rayleigh scattering inside the microresonator \cite{Gorodetsky:00}, a part of the laser radiation is resonantly backscattered (see Fig. \ref{fig:scheme})  to the laser cavity, locking the laser radiation frequency to the frequency of the microresonator mode \cite{Kondratiev:17}. 

It was shown that this system can be described with the nonlinear rate equation (see \cite{Kondratiev:17})
\begin{align} 
\dot A&+\left [\frac{\kappa_{\rm LC}}{2} -\frac{g}{2}(1+i\alpha_g) -i(\omega-\omega_{\rm LC})\right ] A=\kappa_{do}B,
\label{LaserC}
\end{align}
where $\omega_{\rm LC}$ and $\kappa_{\rm LC}$ are the laser cavity eigenfrequency and loss rate, $\kappa_{do}$ is its output mirror coupling rate, $g=g(|A|^2)$ is the laser gain, $\alpha_g$ is the laser medium Henry factor, $\omega$ is the laser generation frequency, $A$ is the laser field slowly-varying complex amplitude, and $B$ is the complex amplitude of the field, reflected from the microresonator. Reflected wave can be described by the following equation
\begin{align}
B(t)=&\sqrt{\Theta}\frac{2i\eta\beta}{(1-i\zeta)^2+\beta^2}A(t-\tau_s) e^{i\omega\tau_s}.
\label{B(t)}
\end{align}
Here the factor $\Theta=S_{\rm LC}/S$, the ratio of the laser aperture area $S_{\rm LC}$ to the final beam area $S$, is introduced to account for the beam refocusing without its power change, which will be needed further. The second multiplier is the microresonator amplitude reflection coefficient \cite{Gorodetsky:00} presented in dimensionless units. It includes the detuning of the laser oscillation frequency $\omega$ from the nearest microresonator eigenfrequency $\zeta=2(\omega-\omega_m)/\kappa_m$ (effective detuning), 
the dimensionless pump coupling coefficient $\eta$ and
the normalized mode-splitting coefficient $\beta$. 
Here $\omega_m$ and $\kappa_m$ are the microresonator mode frequency and the loaded linewidth (loss rate) and
$\tau_s$ is the round-trip time from the laser to the microresonator.

It is convenient to use the tuning curve for analysis of the self-injection locking effect. The curve shows the dependence of the effective frequency detuning $\zeta$ on the detuning of the laser cavity frequency $\omega_{\rm LC}$  from the microresonator eigenfrequency $\xi=2(\omega_{\rm LC}-\omega_m)/\kappa_m$. The tuning curve can be described by the following expression \cite{Kondratiev:17}:
\begin{align}
\label{master}
\xi=\zeta+\tilde\kappa_{do}\frac{4\eta\beta}{\kappa_m}&\frac{2\zeta\cos\bar\psi+(1+\beta^2-\zeta^2)\sin\bar\psi}{(1+\beta^2-\zeta^2)^2+4\zeta^2},\\
\label{masretphase}
&\bar\psi=\psi+\frac{\kappa_m \tau_s}{2}\zeta,
\end{align}
where $\psi=\omega_m \tau_s - \arctan \alpha_g - 3/2\pi$ is the locking phase \cite{Kondratiev:17},  defined by the round-trip time $\tau_s$ from laser to microresonator, by the microresonator resonant frequency $\omega_m$ and the Henry factor $\alpha_g$,
$\tilde\kappa_{do}=\kappa_{do}\sqrt{\Theta}\sqrt{1+\alpha_g^2}$ is the modified coupling rate of the laser cavity and the Henry factor-related laser cavity frequency shift is included into $\omega_{\rm LC}$.
The coupling coefficients also can be expressed in terms of more common coupling rates $\eta = \kappa_c/(\kappa_0 + \kappa_c)$ and $\beta = 2\gamma/(\kappa_0 + \kappa_c)$, where $\kappa_c$ and $2\gamma$ are the pump and forward-backward wave coupling rates and $\kappa_0$ is the intrinsic microresonator loss rate ($\kappa_m=\kappa_c+\kappa_0$). 

\begin{figure}[ht]
\centering
\includegraphics[width=1\linewidth]{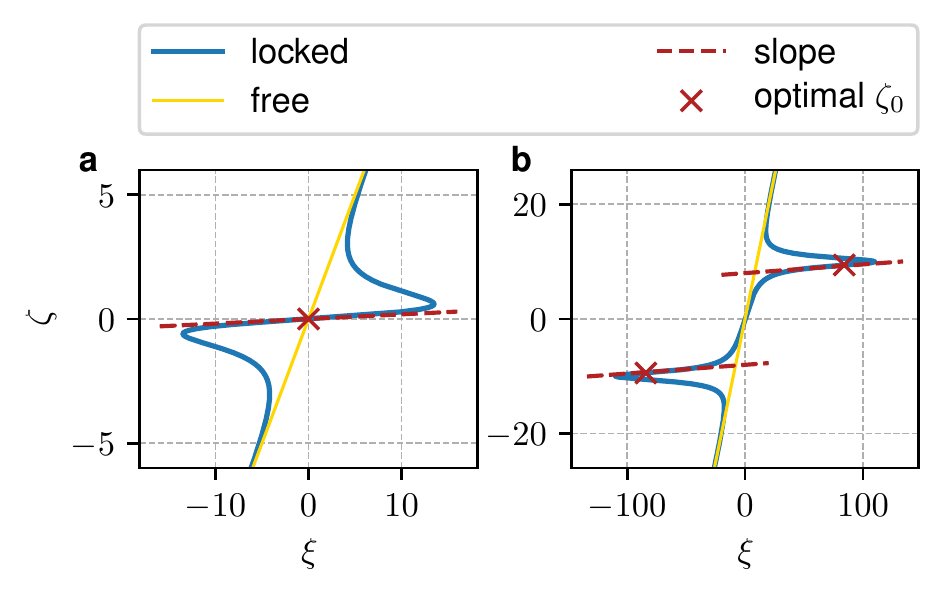}
\caption{The blue lines show the tuning curves for $\beta=0.1$ (panel \textbf{a}) and $\beta=10$ (panel \textbf{b}) in the self-injection locking regime. Here $\psi=0$, $\eta\tilde\kappa_{do}/\kappa_m=50$ and $\kappa_m\tau_s=0.011$. The yellow lines show tuning curves ($\zeta=\xi$) in the free-running regime, the red dashed lines show the slope of locking bands and the red crosses show the optimal points $\zeta=\zeta_0$ \eqref{zeta0}.}
\label{fig:tune}
\end{figure}

The illustrative tuning curves for high and low values of the mode-splitting coefficient $\beta$ are presented in Fig.~\ref{fig:tune}. Note that the tuning curve experiences splitting similar to the resonance splitting at the increased forward-backward wave coupling \cite{Gorodetsky:00}. In this work we show that splitting impacts the self-injection locking process and the stabilization can become worse at larger splitting values.

The both parts in the right hand side of \eqref{masretphase} depend on the feedback round-trip time $\tau_s$. In what follows we consider the $\psi$ to be ``independent'' on $\tau_s$ as the self-injection locking process is periodic on the locking phase and, thus, its absolute value is irrelevant. The scales of $\kappa_m\tau_s$ and $\omega_m\tau_s$ (which is a part of $\psi$) are quite different for high-Q microresonators suggesting to treat these values separately. More formally, the parameter $\psi$ also can be independently tuned with locking mode frequency.

To summarize, one can see that the performance of the laser in the self-injection locking regime is defined by the five main parameters: a) the coupling strength of the forward and backward waves in the cavity $\beta$; b) the locking phase $\psi$ determined by the optical path between the laser and the microresonator and the frequency of the microresonator locking mode; c) the optical round-trip time $\tau_s$ between the laser and the microresonator; d) the laser cavity-microresonator frequency detuning $\xi$, and e) the pump coupling efficiency $\eta$. In what follows we consider effective detuning $\zeta$ instead of the normalized frequency difference between the laser cavity mode and the WGM, $\xi$, since $\xi \ll 1$ in the case of tight injection locking.

To illustrate the importance of these five parameters we simplify the equation \eqref{master} to the classical frequency pulling equation which is valid in the case of the frequency matching between the laser cavity and the microresonator, as well as optimal phase delay between those two ($\cos\bar\psi=1$ and $\omega_{LC} \simeq \omega_m$). Using these approximations we get
\begin{equation} \label{trivial}
    \frac{\omega-\omega_{LC}}{\tilde\kappa_{do}}=-\frac{2\kappa_c}{\kappa_m} \frac{4\beta}{(1+\beta^2)^2} \frac{\omega-\omega_m}{\kappa_m}.
\end{equation}
Eq. \ref{trivial} is analogous to the Adler's formula for master-slave injection locking \cite{Adler:1946,Adler:1973}.
From equation \eqref{trivial} we immediately see that the small backscattering parameter value $\beta \ll 1$, that was considered in the earlier studies involving Fabry-Perot lasers stabilization with an external mirror as well as self-injection locking using WGMRs, is not optimal. It is also evident that the large backscattering is not optimal for the frequency pulling as well. The effect is maximized for the critical coupling $2\kappa_c=\kappa_m$. It is worth noting that the total output light power is nonzero for the critical coupling if the laser beam is not mode matched with WGM ($S_0<S$).

\begin{table*}[htbp]
\begin{center}
\begin{tabular}{|c|c|c|c|c|c|c|}
	\hline
	Sym. & Definition & Values & & Sym. & Definition & Values\\
	\hline
    $K$ & stabilization coefficient: $\partial \xi / \partial \zeta$ &  \numrange{0}{1000} & & $\omega$ & the system generation frequency& $\approx$ 194 THz \\
	\hline
     $\xi$ & normalized LC detuning: $2 (\omega_{\rm LC} - \omega_m) / \kappa_m$ & \numrange{-500}{500} & & $\omega_{\rm LC}$ & laser cavity (LC) frequency &  $\approx$ 194 THz \\
    \hline
    $\zeta$ & normalized generation detuning: $2 (\omega - \omega_{\rm m}) / \kappa_m$& \numrange{-500}{500} & & $\omega_m$ & microresonator mode frequency & $\approx$ 194 THz  \\
    \hline
     $\mu$ & backscattering coefficient: $2\gamma/\kappa_0$ & \numrange{0.01}{10} & & $\kappa_0$ & microresonator intrinsic linewidth &\numrange{0.1}{10} MHz  \\
    \hline
    $\beta$ & normalized backscattering: $2\gamma/\kappa_m$ & \numrange{0.01}{4} & & $\kappa_m$ & microresonator loaded linewidth &\numrange{0.1}{100} MHz \\
    \hline 
     $\eta$& coupling coefficient: $\kappa_c / \kappa_m $ & \numrange{0}{1} & &  $\kappa_c$ & coupling rate ($\kappa_c = \kappa_m - \kappa_0)$ &\numrange{0.1}{100} MHz \\
    \hline
     $\Theta$ & laser aperture -- coupling spot area ratio &  0--1 & & $\gamma$& backscattering rate& \numrange{0.001}{1} MHz  \\
    \hline
    $\psi$ & locking phase: $\psi\approx\omega_m \tau_s$ &  0--$2\pi$ & & $\tau_s$ & round-trip time of feedback &  \numrange{0.01}{0.1} ns \\
    \hline
     $\kappa_0\tau_s$ & normalized round-trip parameter & 0--0.01 & & $\kappa_{do}$ & laser output mirror coupling rate & 10--300 GHz \\
    \hline
\end{tabular}
\caption{Definition of the most important physical parameters describing the self-injection locked laser system and their typical values. The left part of the table includes the dimensionless parameters of the model. The right part of the table contains dimensional parameters to ease the comparison with the experimental values.}
\label{tab:table}
\end{center}
\end{table*}

\section{Laser Linewidth Reduction}

The shot noise limited laser linewidth is reduced proportionally to the square of the stabilization coefficient \cite{Laurent1989,Spano1984} determined by the slope of the tuning curve $K(\eta,\beta,\zeta,\psi)  = \partial \xi/\partial \zeta$. The free-running and locked laser linewidths are related as 
\begin{align}
\delta\omega_{\rm locked}=\frac{\delta\omega_{\rm free}}{K^2}.
\end{align}

A simple formula for the linewidth reduction was obtained in \cite{Kondratiev:17} under conditions of small backscattering $\beta\ll1$, zero locking phase $\psi=0$, resonant tuning $\zeta=0$ and critical coupling $\eta=0.5$. In what follows we perform the 5-parameter ($\psi, \ \zeta,\ \eta,\ \beta, \ \kappa_0 \tau_s$) optimization study of the stabilization coefficient. Taking the derivative of Eq.~\eqref{master} and substituting  $\kappa_m=\kappa_0/(1-\eta)$ in there we get:
\begin{align}
\label{stabilization}
K=&1+4\frac{\tilde\kappa_{do}}{\kappa_0}\eta(1-\eta) \beta \frac{a\cos\bar\psi+b\sin\bar\psi}{((1+\beta^2-\zeta^2)^2+4\zeta^2)^2},\\
a=&-2(3\zeta^4-2(\beta^2-1)\zeta^2-(\beta^2+1)^2)\nonumber\\
  &-\frac{\kappa_m\tau_s}{2}(\zeta^6-(3\beta^2-1)(\zeta^2-\beta^2-1)\zeta^2-(\beta^2+1)^3),\nonumber\\
b=&2\zeta(\zeta^4-(2\zeta^2-\beta^2+3)(\beta^2+1))\nonumber\\
  &-\kappa_m\tau_s\zeta(\zeta^4-2(\beta^2-1)\zeta^2+(\beta^2+1)^2).\nonumber
\end{align}
In contrast with the common knowledge, the increase of the backscattering, described by the parameter $\beta$, does not monotonously enhance the stabilization coefficient \eqref{stabilization}, but leads to its eventual saturation. The optimal selection of the system parameters results in reduction of the laser linewidth by several orders of magnitude. This is the main result of the paper. We consider a few examples of the system optimization.

\subsection{Zero-phase case}

\begin{figure*}[ht]
\centering
\includegraphics[width=0.9\linewidth]{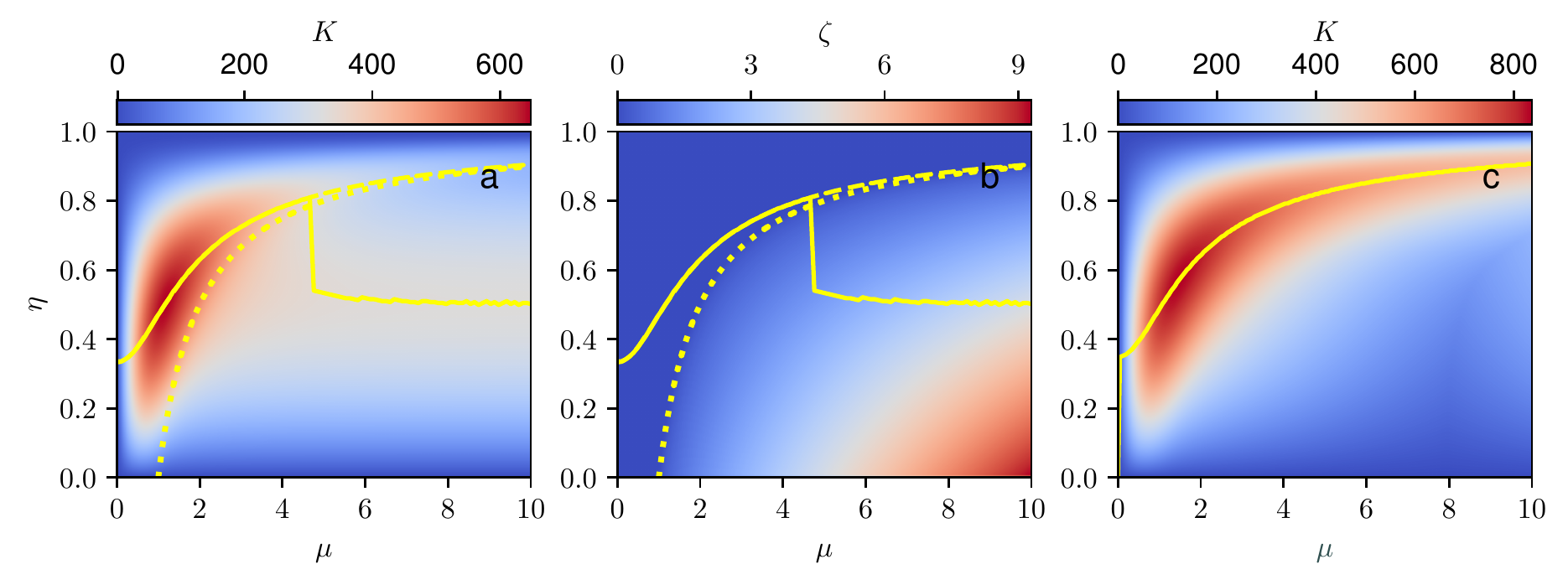}
\caption{The stabilization coefficient $K$ (panel \textbf{a}) and the optimal detuning $\zeta$ (panel \textbf{b}) for $\psi = 0$, $\kappa_0\tau_s=0$, $\frac{\kappa_{\rm do}}{\kappa_0}=1000$. The panel \textbf{c} shows the stabilization coefficient for $\kappa_0\tau_s=0.4$ and the same other parameters. The solid line is numerical maximum with respect to $\eta$, the dashed line is \eqref{eta0} (points near $\mu=1$ omitted), the dotted line corresponds to $\beta=1$.
}
\label{fig:inphase}
\end{figure*}

In this subsection we find the values of the parameters defining the optimal working point of the laser ($\zeta_0$,$\eta_0$,$\beta_0$) for the zero-phase difference $\psi=0$ case. We also show that $\beta$ is not a suitable parameter for the optimization in an experiment and introduce the backscattering coefficient parameter $\mu$ which is an intrinsic parameter of the resonator and does not depend on the setup parameters. We also discuss the impact of the feedback round-trip time $\kappa_0\tau_s$ on the optimal operation point of the system. 

The parameters $\psi=0$ and $\kappa_m \tau_s\ll1$ are selected as the most common and illustrative. In this case the resonance curve of the locking mode (for a laser diode it coincides with the Light-Current (LI) curve), which can be observed while the laser frequency is broadly scanned in and out of the locking range, has a nearly rectangular shape \cite{Kondratiev:17,photonics5040043}. The phase that the light accumulates while traveling between the laser and the resonator $\psi$ can be adjusted either by tuning the laser-microresonator distance or by choosing the locking mode with appropriate frequency, in accordance with $\psi=\omega_m \tau_s - \arctan \alpha_g - 3/2\pi$.  Different modes can have different backscattering $\beta$ \cite{Zhu2010,Li:12}. Practically, a mode with desired $\psi$ and $\beta$ can be selected by analyzing the LI curve (see Appendix \ref{sec:exampleOpt}).

To optimize the laser performance we look for the effective detuning $\zeta_0$ that maximizes the stabilization coefficient $K$. The expression $\left.\partial K(\zeta,\psi=0,\eta,\beta)/\partial\zeta\right|_{\zeta_0}=0$ results in a $\zeta$-multiplied bi-cubic characteristic equation, dependent on $\beta$ only. Solving it we obtain for different $\beta$ values
\begin{align}
    \label{zeta0}
    \zeta_0=
    \begin{cases}
        0, &(\beta\leq1)\\
        \sqrt{\frac{3}{5}(\beta^2-1)}, &(\beta\in(1;1.48))\\
        \beta - \frac{1}{\sqrt{3}}+... &(\beta>1.48)
    \end{cases}
\end{align}
This expression has a simple physical meaning. The linear interaction of the counter-propagating waves leads to the resonance splitting \cite{Gorodetsky:00}. The splitting value is approximately equal to $\beta\kappa_m$ (see Fig. \ref{fig:scheme}). The locking band splits into two for the large $\beta$ [see Fig. \ref{fig:tune}(b), the crosses mark the points $\zeta=\pm(\beta-1/\sqrt{3})$ and the tips of the peaks are close to $\zeta=\pm\beta$].  

We can rewrite the expression for the stabilization coefficient as a product of two parts, where one part depends solely on $\eta$ and the other part depends solely on $\beta$. Substituting Eq.~\eqref{zeta0} into Eq.~\eqref{stabilization} we obtain:
\begin{align}
\label{psi=0}
K(\eta,\beta,&\zeta_0,0)=1+8\frac{\tilde\kappa_{do}}{\kappa_0}\eta(1-\eta) \beta \times& \nonumber\\
&\times
\begin{cases}
    \frac{1}{(1+\beta^2)^2},\ &(\beta\leq1)\\
    \frac{25}{4}\frac{7\beta^4+11\beta^2+7}{(\beta^4+23\beta^2+1)^2},\ &(\beta\in(1;1.48))\\
    \frac{27}{32}\beta\frac{\beta^2\sqrt{3}-2\beta+\sqrt{3}}{(3\beta^2-\beta\sqrt{3}+1)^2}.\ &(\beta>1.48)
\end{cases}
\end{align}
The slope has a maximum inside its validity range if $\beta\leq1$. The maximum of the stabilization coefficient is
\begin{align}
    \label{global0}
    K_{\rm max}^0\approx&\frac{3\sqrt{3}}{8}\frac{\tilde\kappa_{do}}{\kappa_0}+1,\\
    \eta_{\rm max}^0=1/2,& \ \beta_{\rm max}^0=3^{-1/2},\ \psi_{\rm max}^0=0\nonumber
\end{align}
In what follows we show that the maximum can be increased for the case of $\kappa_m\tau_s>0$.

For large $\beta$ the stabilization coefficient becomes independent on $\beta$ and follows the coupling constant
\begin{align}
\label{Kbinfp0}
K_{\beta\rightarrow\infty}^0\approx\frac{3\sqrt{3}}{4}\eta(1-\eta)\frac{\tilde\kappa_{do}}{\kappa_0}+1.
\end{align}
It optimizes at $\eta_{\beta\rightarrow\infty}^0=1/2$ reaching value $K_{\beta\rightarrow\infty}^0\rightarrow(3\sqrt{3}/16)(\tilde\kappa_{do}/\kappa_0)+1$. This is  twice smaller than the maximum stabilization \eqref{global0}.

Let us summarize the results of this subsection. We found that the optimal detuning is nonzero for strong backscattering. The stabilization coefficient is not the largest in this case, though. This high-$\beta$ limit corresponds to $K_{\beta \gg 1}$ two times lower than the low-beta level $K_{\rm max}^0$.

\subsection{Zero-phase and fixed backscattering case}

The forward-backward wave coupling rate $2\gamma$ is usually fixed in an experiment. It is possible to optimize the laser stabilization by varying the values of the detuning $\zeta$, locking phase $\psi$ and the coupling coefficient $\eta$ only. Since $\beta$ depends on the total bandwidth $\kappa_m$ and, thus, on $\eta$, it can not be used as an independent optimization parameter.  

We introduce another parameter $\mu = 2\gamma/\kappa_0$ ($\beta = \mu (1 - \eta)$) that is a constant for a given resonator. Using this notation we perform a full parametric optimization of the stabilization coefficient. Figure \ref{fig:inphase}(a) shows the results of the numerical optimization for zero phase $\psi=0$ and optimal frequency detuning \eqref{zeta0}. It can be seen that the critical coupling is optimal for the short laser-microresonator distance  ($\kappa_0\tau_s<0.1$) and for the large backscattering $\beta\geq1$. 

The dependence of the optimal pump coupling coefficient $\eta$ on the normalized forward-backward coupling rate $\mu$ is shown by the solid line. While increasing $\mu$ we should increase the load to keep $\beta<1$, preventing the resonance splitting ($\beta=1$ is shown in Fig. \ref{fig:inphase} with the dotted line). This is also clearly seen in the map of the optimal detuning (see Fig. \ref{fig:inphase}(b)). At some point (at $\mu \approx 5$ for the considered parameters) the detuning increase is not advantageous any longer and the critical coupling becomes optimal. 

The approximation of the first part of the $\eta(\mu)$ curve can be found using expression  \eqref{psi=0} for $\beta<1$ and the optimization of $K$ with respect of $\eta$. Substituting $\beta=\mu(1-\eta)$ into \eqref{psi=0} and differentiating $K$ with respect of $\eta$ results in a cubic characteristic equation \eqref{eta0eq} (see Appendix \ref{App1}).
This equation is quadratic with respect to $\mu$. It has an exact solution (see Appendix \ref{App1}, \eqref{muopt0}), but it is more illustrative to solve the equation with respect to $\eta$. 

There are three roots for $\eta$, only one of which belongs to the $[0;1]$ region. We find an asymptotic solution of this root 
\begin{align}
\label{eta0}
\eta_{0}&\approx
\begin{cases}
    \frac{1}{3}+\left(\frac{2}{3}\right)^4\mu^2-\left(\frac{2}{3}\right)^6\mu^4, &\mu\ll1\\
    1-\frac{1}{\mu}+\frac{1}{2\mu^2}. &\mu\gg1
\end{cases}
\end{align}
This solution still represents a local maximum for $\mu>5$, but its value decreases. The maximum stays constant \eqref{Kbinfp0} at the critical coupling regime and becomes the main one (see also the curves ``$\psi=0$'' and ``2-branch'' in Fig. \ref{fig:max_K}).

We checked the analytical results and studied the nonzero $\kappa_m\tau_s$ case numerically using \eqref{stabilization}. For a relatively large distance between the laser and the microresonator  ($\kappa_m\tau_s>0.1$) the overcoupled regime is optimal at larger forward-backward coupling coefficient values (see Fig. \ref{fig:inphase}(c)). For example,  $\kappa_0\tau_s=0.1$ corresponds to $l_{\rm cr}\approx0.1Q\lambda/(2\pi)=247$~mm for $Q=10^7$. The absolute value of $K$ grows with $\kappa_0\tau_s$ (see Fig. \ref{fig:inphase}(c) and curve ``$\kappa_0\tau_s=0.4$'' in Fig. \ref{fig:max_K}).

\subsection{Optimal locking phase}
\label{section:optimla_locking_phase}

In this subsection we show that the maximum of the low-$\beta$ limit found in the previous subsection does not change with the locking phase $\psi$ optimization for a wide range of the feedback round-trip time values, while the high-$\beta$ limit increases further. We also derive corrections related to a larger $\kappa_m\tau_s$ parameter.

Equation \eqref{stabilization} can be optimized with respect of $\psi$ (see Appendix \ref{App2}). For the optimal phase we get
\begin{align}
\label{psi_0}
\psi_{\rm opt}=\alpha-\frac{\kappa_m\tau_s}{2}\zeta+\pi n, 
\end{align}
where $\sin\alpha=b/(a^2+b^2)^{1/2}$. Equation \eqref{stabilization}  is simplified as
\begin{align}
\label{stabilizationpsi}
K&=1+4\frac{\tilde\kappa_{do}}{\kappa_0}\eta(1-\eta) \beta \frac{\sqrt{a^2+b^2}}{((1+\beta^2-\zeta^2)^2+4\zeta^2)^2}.  
\end{align}
The derivative of \eqref{stabilizationpsi} with respect of $\zeta$ is a bi-cubic equation, multiplied by $\zeta$ \eqref{zetaopteqass}. The equation is turned to a bi-quadratic one if $\kappa_m\tau_s$ is neglected (small round trip time is expected), resulting in
\begin{align}
\label{zetaopt}
\zeta_{\rm opt}=
\begin{cases}
    0, &\beta\leq \beta_{\rm cr}\\
    \sqrt{\frac{\beta^2}{3}-1+\frac{2}{3}\sqrt{\beta^2(\beta^2+3)}}, &\beta>\beta_{\rm cr}
\end{cases}
\end{align}
where $\beta_{\rm cr}=\sqrt{2\sqrt{3}-3}\approx 0.68$ is the value of $\beta$ at which the nonzero root becomes real. We selected the positive sign for $\zeta_{\rm opt}$, but the parameter also can be negative. This sign switch changes the sign of $\psi_{\rm opt}$ (see \eqref{psi_0}). For small $\beta$ ($\beta\leq \beta_{\rm cr}$), and, thus, small $\mu$, we get the same $\zeta_{\rm opt}=0$ (see also  Fig. \ref{fig:full}(b)) as for the zero-phase case. This critical value $\beta_{\rm cr}$ increases with the round-trip time and pump coupling coefficient ($\kappa_m\tau_s$) but always stays less than unity (see Appendix \ref{App2}). This dependence can be approximated as follows
\begin{align}
\label{betacr}
\beta_{\rm cr}=1-2\frac{1-\sqrt{2\sqrt{3}-3}}{\kappa_m\tau_s(1-\sqrt{2\sqrt{3}-3})+2}.
\end{align}
Substituting $\zeta=0$ to the expression \eqref{stabilization} we get $b=0$. Substituting this result into the expression for the optimal phase \eqref{psi_0} we find that $\psi_{\rm opt}=0$ (see  Fig. \ref{fig:full}(c)). Therefore, the optimum is universal for an arbitrary $\kappa_m\tau_s$ value. The optimum for the case of small $\beta$ we found in the previous section is absolute.

At the next step we use $\kappa_m\tau_s=\kappa_0\tau_s/(1-\eta)$ and evaluate the global maximum taking $\kappa_0\tau_s \ne 0$ into account. We substitute \eqref{zetaopt} into \eqref{stabilizationpsi}.  It can be shown numerically that there is no localized maximum for $K$ in the high-beta regime. For the case of small $\beta$ we get
\begin{align}
\label{Kzopt}
K|_{\beta<\beta_{\rm cr}}=&1+2\frac{\tilde\kappa_{do}}{\kappa_0}\eta \beta     \frac{4(1-\eta)+\kappa_0\tau_s(1+\beta^2)}{(1+\beta^2)^2}.
\end{align}
Performing optimization, we obtain the final expressions for the global maximum
\begin{align}
\label{global}
\beta_{\rm max}&=\sqrt{\frac{6-\kappa_0\tau_s-2\sqrt{9-4\kappa_0\tau_s}}{\kappa_0\tau_s}}\nonumber,\\
\eta_{\rm max}&=\frac{5}{4}-\frac{1}{4}\sqrt{9-4\kappa_0\tau_s},\\
K_{\rm max}&=1+\frac{\tilde\kappa_{do}}{\kappa_0}\frac{(3+2\kappa_0\tau_s+\sqrt{9-4\kappa_0\tau_s})^2}{32}\beta_{\rm max}.\nonumber
\end{align}
{\em This expression describing the global maximum of the coefficient $K$ as well as the corresponding optimal parameters enabling the maximum is the main result of the paper.} Realizing the indicated parameter values in the experiment ensures the optimal performance of the device. The optimal $\beta$ is mostly determined by the microresonator material and its mechanical processing and usually cannot be tuned (the tuning is still possible if a micro-object is placed within the evanescent field of the mode\cite{Mazzei:07,Zhu2010}), other parameters can be tuned in real time. 

Let us consider a few special cases allowing the simplification of the expressions\eqref{global}. For the case of high-Q microresonators and the small distances $\kappa_0\tau_s \ll 1$, we expand these expressions into series with respect of $\kappa_0\tau_s$ and find that they reduce to simple corrections to formula  \eqref{global0}
\begin{eqnarray}
\label{global_expansion}
\beta_{\rm max}\approx\beta_{\rm max}^0+\frac{2\sqrt{3}}{27}\kappa_0\tau_s \nonumber, \\ \eta_{\rm max}\approx\eta_{\rm max}^0+\frac{1}{6}\kappa_0\tau_s, \\ K_{\rm max}\approx K_{\rm max}^0+\frac{\tilde\kappa_{do}}{\kappa_0}\frac{\sqrt{3}}{4}\kappa_0\tau_s.\nonumber 
\end{eqnarray}
All three quantities grow with $\kappa_0\tau_s$ increase.  

According to \eqref{betacr}, the border of the low-$\beta$ region is also shifted so that the global maximum lays almost at the border in the low-$\beta$ region (see also Fig. \ref{fig:betaregion}). The maximum is still described by \eqref{global} until the $\kappa_0\tau_s=2$, where $\eta_{\rm max}$ becomes greater than 1. This naturally means that the stabilization coefficient $K(\eta,\beta,\zeta_{\rm opt},\psi_{\rm opt})$ monotonically increases with $\eta$. The coefficient $K$ ceases to have any extremum features with respect to $\eta$. In the overcoupled state ($\eta\approx1$) $K$ saturates with $\beta$.

\begin{figure}[ht]
\centering
\includegraphics[width=.99\linewidth]{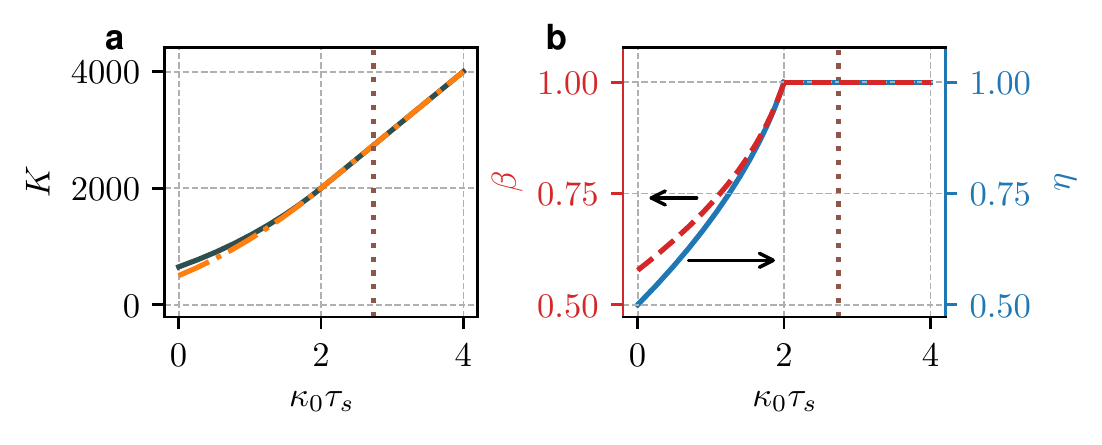}
\caption{\textbf{a)} High-$\beta$ limit of the stabilization coefficient  \eqref{Kbinf} with \eqref{etaopt} (orange dash-dot line) and its global maximum value \eqref{global} (dark-slate-grey solid line) for different round-trip time. \textbf{b)} The optimal parameters $\eta$ (right axis, solid line) and $\beta$ (left axis, dashed line) for the different regimes of $\kappa_0 \tau_s$. The vertical brown dotted line on both panels ($\kappa_0 \tau_s = 2.74$) shows the boundary of redundant fringe appearance \cite{Kondratiev:17} ($\kappa_0 \tau_s < 2.74$ -- non-fringe region). The parameters $\zeta = 0$ and $\psi = 0 $ are optimal in all the regimes of $\kappa \tau_s$. }
\label{fig:K_max_its_parameters}
\end{figure}

To describe this regime, we can use asymptotic expression of \eqref{stabilizationpsi} and \eqref{zetaopt} for large $\beta$. For the stabilization coefficient we get
\begin{align}
\label{Kbinf}
K_{\beta\rightarrow\infty}&\approx\frac{\tilde\kappa_{do}}{\kappa_0}\eta(\kappa_0\tau_s+2(1-\eta))+1.
\end{align}
This is similar to the zero-phase case expression \eqref{Kbinfp0} (but 1.5 times larger for low $\kappa_0 \tau_s$). The result also indicates the transition of the optimal regime from the critically coupling to the overcoupled cavity for large round-trip times.

Finally, for the saturated stabilization coefficient at the condition of large $\kappa_0\tau_s$ we obtain
\begin{align}
\label{global-high}
\beta_{\kappa_0\tau_s>2}>&1,\nonumber\\
\eta_{\kappa_0\tau_s>2}=&1,\\
K_{\kappa_0\tau_s>2}=&\frac{\tilde\kappa_{do}}{\kappa_0}\kappa_0\tau_s+1\nonumber.
\end{align}
The values of the optimal stabilization coefficient and corresponding optimal parameters are shown in Fig. \ref{fig:K_max_its_parameters}. The $\kappa_0\tau_s=2$ corresponds to the total optical distance $l_{\rm cr}\approx2Q\lambda/(2\pi)\approx5$~m for $Q=10^7$. We also note that the absolute value of $K$ grows proportionally to $\kappa_0\tau_s$.

Using  \eqref{stabilizationpsi} with $\beta=\mu(1-\eta)$ we derive an expression for the optimal pump coupling as the function of the backscattering (see Appendix \ref{App3}). The exact solution can be obtained for $\mu(\eta)$ \eqref{muopt}, but the solution for $\eta$ is more illustrative.
Assuming small $\mu$ in \eqref{Kzopt} and expanding the result into series with respect of $\kappa_0\tau_s$ we find the correction to the low-$\beta$ case.  The corrections for the optimal coupling coefficient $\eta_{\rm opt}$ for high-$\beta$ region can be also found using \eqref{Kbinf}. As the result we obtain
\begin{align}
\label{etaopt}
\eta_{\rm opt}&\approx
\begin{cases}
    \frac{1}{3}+\left(\frac{2}{3}\right)^4\mu^2-\left(\frac{2}{3}\right)^6\mu^4+\frac{\kappa_0\tau_s}{24}, \!&\beta<\beta_{\rm cr}\\
    \frac{1}{2}+\frac{\kappa_0\tau_s}{4}. \!&\beta\gg\beta_{\rm cr}
\end{cases}
\end{align}

\begin{figure*}[ht]
\centering
\includegraphics[width=0.9\linewidth]{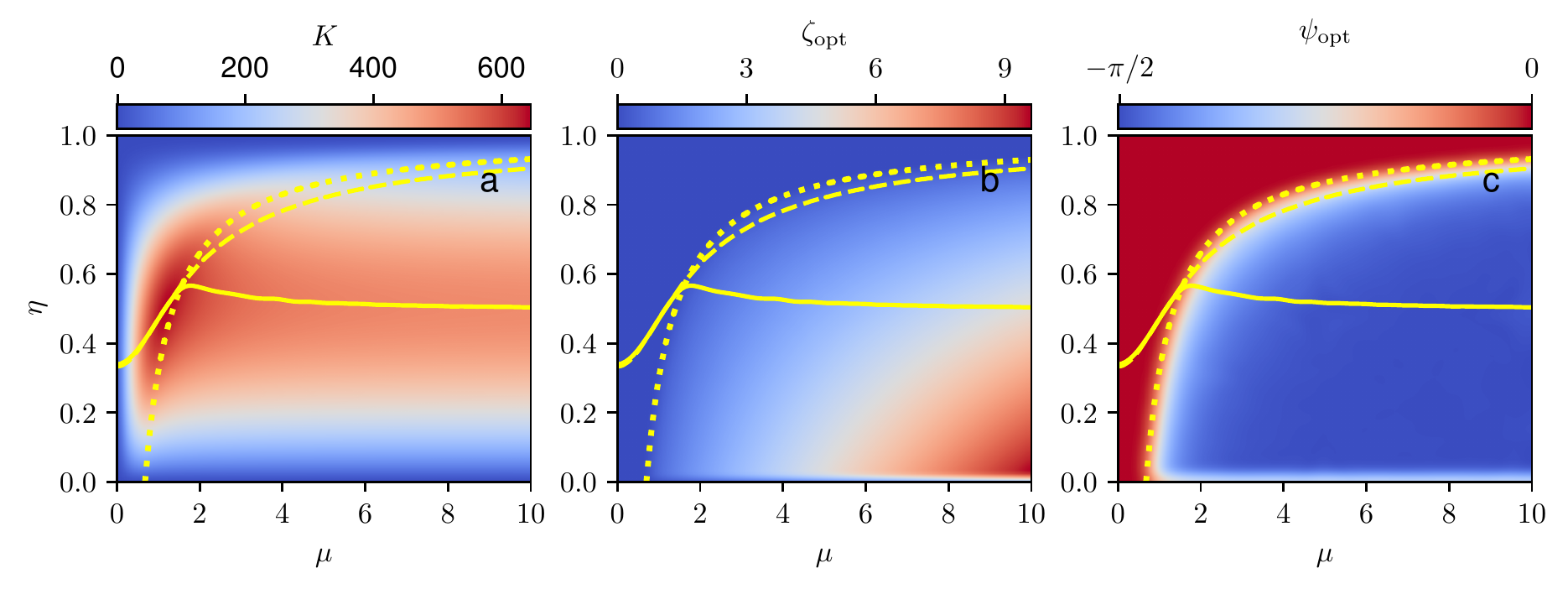}
\caption{The stabilization coefficient $K$ (panel \textbf{a}), the optimal detuning $\zeta_{\rm opt}$ (panel \textbf{b}) and optimal $\psi_{\rm opt}$ (panel \textbf{c}) for $\kappa_0\tau_s=0$, $\frac{\kappa_{\rm do}}{\kappa_0}=1000$. The solid line is numerical maximum with respect to $\eta$, the dashed line is \eqref{eta0} (points near $\mu=1$ omitted), the dotted line corresponds to $\beta=\beta_{\rm cr} \approx0.68$. 
}
\label{fig:full}
\end{figure*}

\begin{figure}[ht]
\centering
\includegraphics[width=0.9\linewidth]{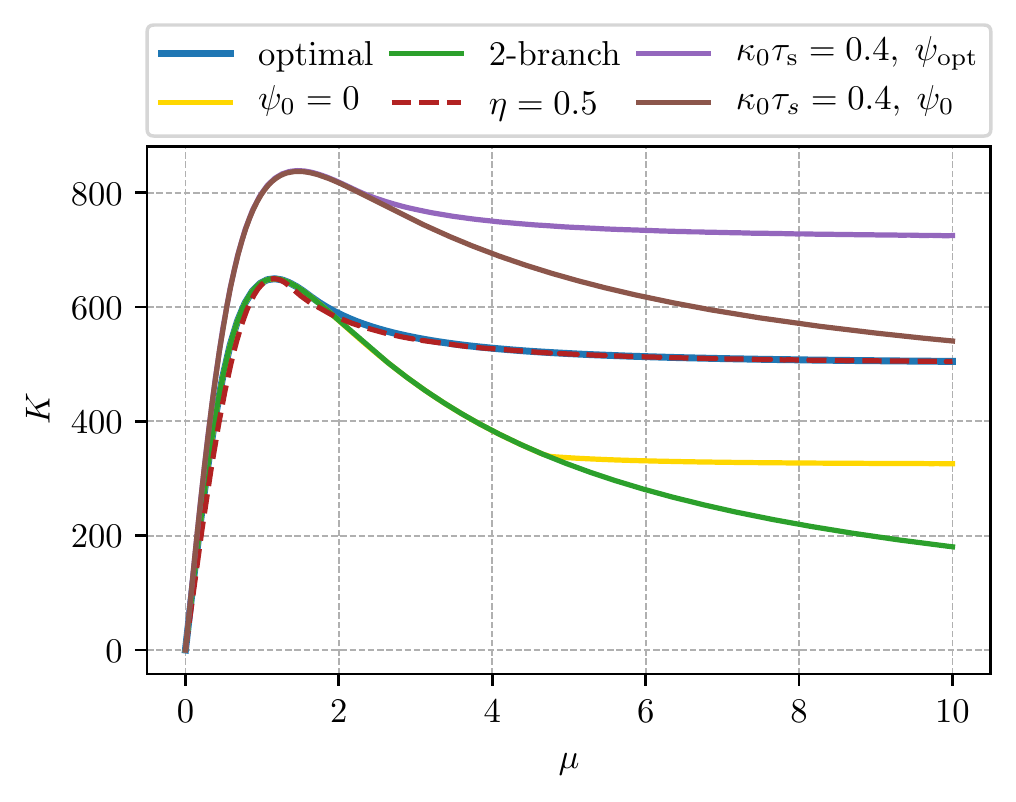}
\caption{The maximum frequency stabilization coefficient depending on the resonator parameter $\mu$ for the different self-injection locking parameters regime. Lines correspond to: \textbf{Blue}: the optimal state ($\psi_{\rm opt} , \ \zeta_{\rm opt} , \ \eta_{\rm opt}$);
\textbf{Green}: the second branch of fixed phase $\psi = 0$ and optimal ($\zeta_0 , \ \eta_0$), \eqref{eta0};
\textbf{Yellow}: fixed phase $\psi=0$ and optimal ($\zeta_{0} , \ \eta_{\rm opt}$); 
\textbf{Red-dashed}: critical coupling $\eta = 0.5$ and optimal ($\psi, \ \zeta$); 
\textbf{Purple}: $\kappa_0 \tau_s = 0.4$ for optimal parameters ($\psi_{\rm opt} , \ \zeta_{\rm opt} , \ \eta_{\rm opt}$);
\textbf{Brown}: $\kappa_0 \tau_s = 0.4$ for fixed phase $\psi=0$ and ($\zeta_{0} , \ \eta_{0}$).
}
\label{fig:max_K}
\end{figure}

\begin{figure*}[ht]
\centering
\includegraphics[width=0.9\linewidth]{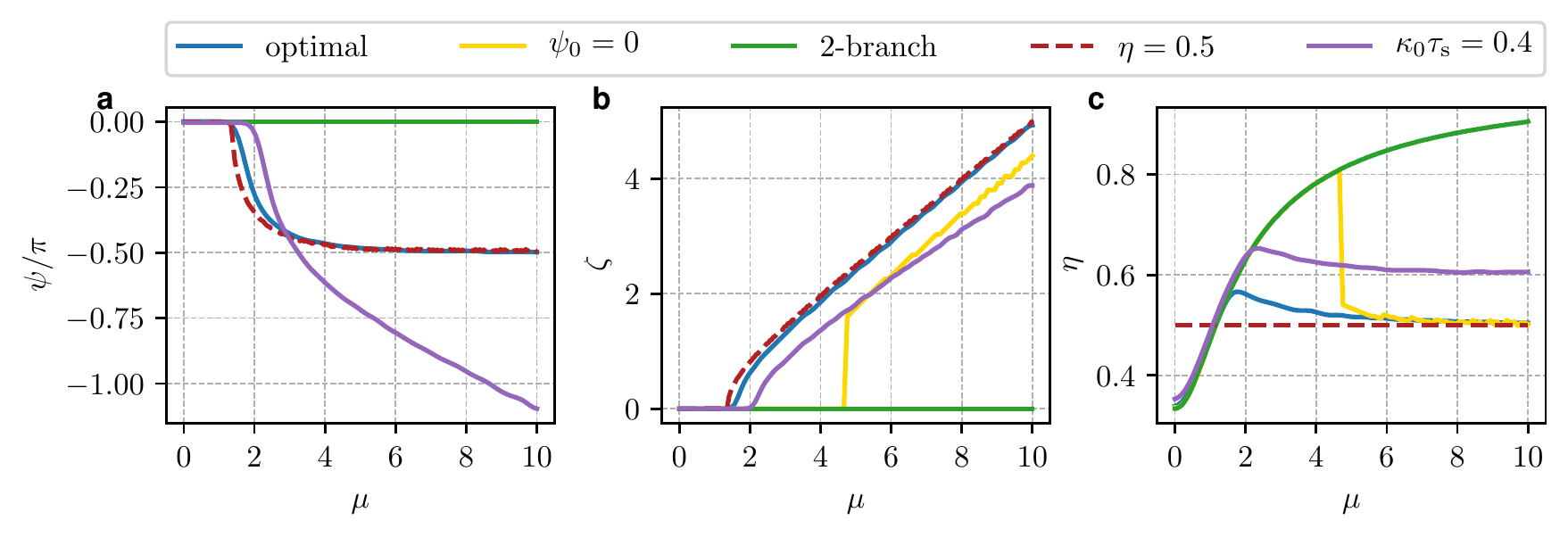}
\caption{ a) The optimal phase shift between the laser and the cavity; b) the optimal tuning of the laser radiation frequency; c) the optimal coupling efficiency. Lines correspond to: \textbf{Blue}: the optimal state ($\psi_{\rm opt} , \ \zeta_{\rm opt} , \ \eta_{\rm opt}$);
\textbf{Green}: the second branch of fixed phase $\psi = 0$ and the optimal ($\zeta_0 , \ \eta_0$),\eqref{eta0};
\textbf{Yellow}: fixed phase $\psi=0$ and the optimal ($\zeta_{0} , \ \eta_{0}$); 
\textbf{Red-dashed}: critical coupling $\eta = 0.5$ and the optimal ($\psi, \ \zeta$); 
\textbf{Purple}: $\kappa_0 \tau_s = 0.4$ for the optimal parameters ($\psi_{\rm opt} , \ \zeta_{\rm opt} , \ \eta_{\rm opt}$);
\textbf{Brown}: $\kappa_0 \tau_s = 0.4$ for fixed phase $\psi=0$ and ($\zeta_{0} , \ \eta_{0}$).
}
\label{fig:optimal_phi_z}
\end{figure*}

The above calculations indicate that the optimal stabilization coefficient increases with the laser-microresonator distance. However, one should not increase the distance uncontrollably as the increase is responsible for the decrease of the laser signal quality and also produces redundant metastable fringes on the tuning curve \cite{Kondratiev:17}. The criterion of the stable operation was approximated as $\kappa_m\tau_s<9.4(8\eta(1-\eta)\beta\tilde\kappa_{do}/\kappa_0)^{-0.36}$. For the maximum point \eqref{global} this criterion is simplified to $\kappa_0\tau_s>2.74$ (see Fig. \ref{fig:K_max_its_parameters}). Furthermore,  $\kappa_0\tau_s> 1$ will lead to increase the mechanical instability noise. These considerations are subject to separate experimental study.

The map of the stabilization coefficient \eqref{stabilizationpsi} under the conditions of the optimal detuning \eqref{zetaopt} and locking phase \eqref{psi_0} is shown in Fig. \ref{fig:full}(a) for different combinations of $\eta$ and $\mu$. The maps of optimal detuning and optimal phase are shown in Figs. \ref{fig:full}(b) and \ref{fig:full}(c). Zero phase ($\psi=0$) is an exact optimum for $\beta<0.68$ (see Fig. \ref{fig:full}(c)), which is connected to the optimal condition $\zeta=0$ (see Fig. \ref{fig:full}(b)). Since the optimal value of $\beta$ found earlier for the zero-phase case $\beta_{\rm max}=3^{-1/2}$ is less than $0.68$, the maximum stabilization coefficient value for the zero locking phase is a global maximum.

The critical coupling $\eta=0.5$, considered in \cite{Kondratiev:17}, is very close to the optimal $\eta(\mu)$ line (see Fig.~\ref{fig:full}a). It is interesting to fix $\eta=0.5$ and perform the optimization with respect of the other parameters. 
Inserting \eqref{etaopt} for small $\mu$ into \eqref{Kzopt} we find: 
\begin{equation}
K_{\rm \mu \rightarrow 0} = 1 + \frac{32}{27} \mu \frac{\tilde\kappa_{do}}{\kappa_0} \left ( 1+\frac{3}{8}\kappa_0 \tau_s \right). 
\end{equation}
For the case of the critical coupling ($\eta = 0.5$) we have a less steep dependence 
\begin{equation}
K_{\rm \mu \rightarrow 0}  = 1  + \mu \frac{\tilde\kappa_{do}}{\kappa_0} \left ( 1 + \frac{1}{2}\kappa_0 \tau_s \right). 
\end{equation}
The difference is quite small. It is only 1.18 times.

The maximum stabilization coefficient is usually reached by selecting modes with an optimal phase shift $\psi$, an optimal coupling $\eta$, as well as an optimal detuning of the laser emission frequency from the resonance $\zeta$ for a given $\mu$. In Fig. \ref{fig:max_K} we present the dependence of $K$ on the resonator parameter $\mu$ for the most interesting regimes. We also show the corresponding optimum values $\psi$, $\eta$ and $\zeta$ in Fig. \ref{fig:optimal_phi_z}. In all the considered regimes the maximum value of the stabilization coefficient is reached at $\mu = \beta_{\rm max}/(1-\eta_{\rm max})<2$. At larger values of $\mu$ it either saturates or monotonously decreases because of the mode splitting.  Active tuning of $\psi$ and $\zeta$ is necessary at $\mu > 2$ to reach the maximum value of the stabilization coefficient [see Fig. \ref{fig:optimal_phi_z} (a) and (b)]. This is hard to do experimentally. Thus, it is desirable to select a mode with $\mu \leq 2$, which corresponds to a 'semi-split' mode (see Fig. 2(d) in \cite{Raja2019}).

According to our model the optimization of the self-injection locking could result in a significant reduction of the laser linewidth if compared with the best experimental results. For example, a diode laser linewidth reduction from 2 MHz in the free-running regime to sub-100 Hz in the locked regime was demonstrated in \cite{Liang2015}. The linewidth reduction in the case of the optimal parameters $\eta_{\rm opt}(\mu), \ \psi_{\rm opt}(\mu)$ and $\zeta_{\rm opt}(\mu)$ for $\mu = 3$ can be improved by $15$ times, which is at least an order of magnitude better than the result obtained for the non-optimal coupling. Furthermore, if the mode of the resonator is optimally selected ($\mu = 1.16$), the linewidth reduction can be improved by $94$ times (see section \ref{sec:exampleOpt} for details). 

To summarize, in this subsection we have found the global maximum for the locking coefficient $K$. The global optimum is situated in the vicinity of
$\beta=3^{-1/2}$, $\eta=1/2$, $\zeta=0$, $\psi=0$ and grows linearly with the value of $\kappa_0\tau_s$.

\section{Discussion}

We showed nonmonotonic saturation of the stabilization coefficient $K$ with respect to the backscattering (see Fig. \ref{fig:max_K}). The maximum value of the stabilization coefficient is reached at {$\beta\in[3^{-1/2};1]$}, which determines the optimal 'semi-split' mode. The saturation happens due to the formation of the doublet backscattering resonance because of  the counterpropagating modes in the microresonator. Microresonator modes with high backscattering rate require the laser frequency to be tuned to the inner slope of the doublet backscattering resonance to achieve the highest stabilization coefficient (see Fig. \ref{fig:optimal_phi_z}). In the following we will consider an exemplary case with recommendations for the optimization implementation and limitations on the model due to the nonlinear effects.

\subsection{Optimal regime realization}

\label{sec:exampleOpt}
In this section we describe the implementation process of the proposed optimization model using the experimental data from \cite{Liang2015}. The experimental parameters are: unloaded-Q $\sim 6 \times 10^9$, loaded-Q $\sim 6 \times 10^8$, mode-splitting value of the order of 100~kHz, which corresponds to $\mu \sim 3$ and $\eta \approx 0.91$. The normalized backscattering is evaluated to be $\beta=0.27$, indicating the low-scattering regime. According to the optimization map in Fig. \ref{fig:full}(b,c) the optimal values of phase and detuning are zero. 

\begin{figure}[ht]
\centering
\includegraphics[width=0.99\linewidth]{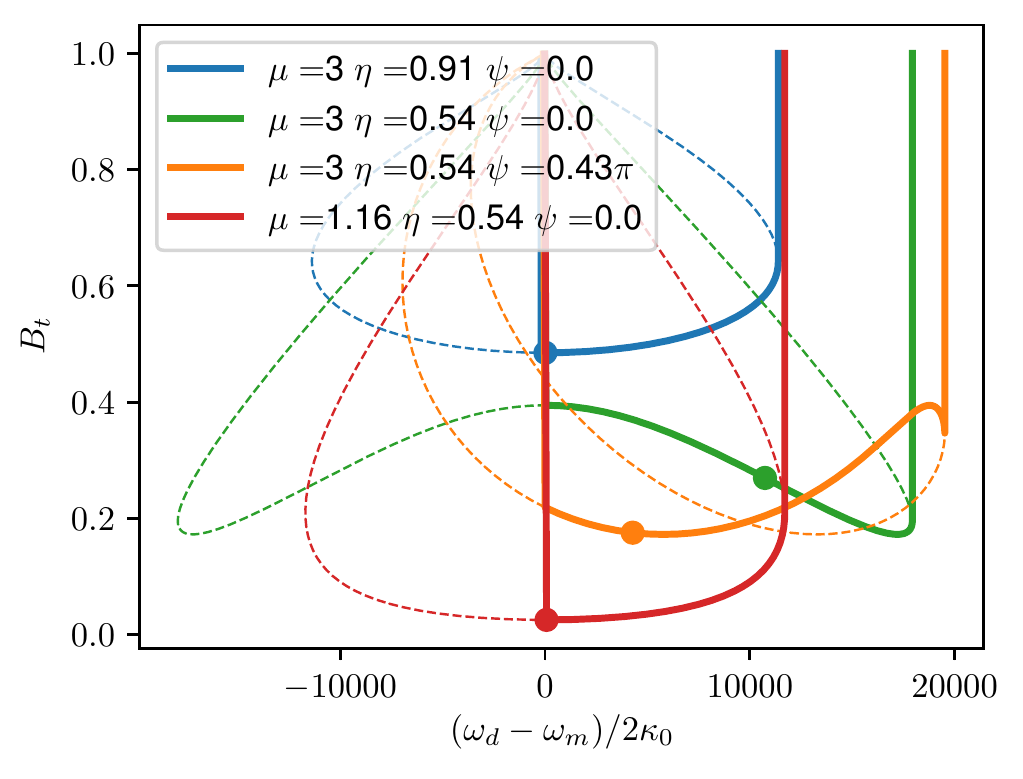}
\caption{Numerically obtained the transmission resonance curves \eqref{Transmission} for the parameters taken from \cite{Liang2015}. The parameter $\zeta(\xi)$ was evaluated using \eqref{master} (the dashed lines), and corresponding LI curves were evaluated for the increasing frequency (the solid lines). The points on the transmission curves mark the optimal detuning values.}
\label{exampleLI}
\end{figure}

We assume that the detuning and phase in the experiment were close to the optimal  $\zeta = 0$ and $\psi = 0$  which provides the typical close-to-rectangular LI curve (see the blue curve in Fig. \ref{exampleLI}). The optimal detuning is marked with a point. The theoretical LI-curve is obtained from the transmission resonance curve, given by \cite{Gorodetsky:00}:
\begin{align}
\label{Transmission}
B_t\approx B_{\rm in}\left(1-2\eta\frac{(1-i\zeta)}{(1-i\zeta)^2+\beta^2}\right)
\end{align}
Only a part of the \eqref{Transmission} curve is accessed during monotonic sweep of the laser frequency \cite{Kondratiev:17} and constitute the LI curve. The figure \ref{exampleLI} shows the resonance curves \eqref{Transmission} with dashed lines and LI curves while increasing frequency (current decrease) with solid lines. Note that for high enough quality factor of the microresonator the locking region entrance point is close to zero detuning, which also helps to tune to the optimum. Another important point is that in the real experiment the current change also changes the laser output power and thus the LI curve can be tilted $B_t'=B_t+(\partial B_{\rm in}/\partial I)(\partial I/\partial \xi)\xi$, where $I$ stands for the current.

According to Fig. \ref{fig:full}(a) the laser system can be further optimized with respect of $\eta$. Selecting better parameters according to the yellow line in Fig. \ref{fig:full}(b,c) we obtain $\eta=0.54$,
$\zeta=-1.3$ (corresponds to $\xi=8.97$),
$\psi=0.43\pi$ . The corresponding linewidth can be improved by $K^2_{\eta=0.54,\mu =3} / K^2_{\eta=0.91,\mu =3} \approx 15$ times, which is an order of magnitude better than in the experiment. We need to switch the signs of $\zeta$ and $\psi$ for the optimum to occur in the forward scan. Decreasing the pump coupling to the optimal value we set it close to the critical coupling, thus transmittance will be reduced \cite{Gorodetsky:00}. At the same time $\beta$ grows to 1.38 and the locked region width grows. Both can be seen in the corresponding LI curve (see green curve in Fig. \ref{exampleLI}). As $\beta$ is now greater then the critical value \eqref{betacr} the optimal phase becomes nonzero. The correct phase also can be controlled using the transmittance resonance form (see the difference between the green and orange curves in Fig. \ref{exampleLI}). If the laser-microresonator distance is unchangeable in a particular setup, the phase can be tuned by switching the operational mode. This step, however, can modify the scattering coefficient $\mu$ as well.

If the mode of the resonator is optimally selected so that $\mu = 1.15$, the linewidth reduction improves by $K^2_{\eta=0.5,\mu =1.15} / K^2_{\eta=0.91,\mu =3} \approx 94$. The LI curve analysis also suggests the optimal mode, that should have the particular shape (see Fig. \ref{exampleLI}, red curve).

In general we can elaborate the following optimization recommendations based on our theoretical model.
\begin{enumerate}
    \item If we can estimate $\mu$, we select a mode with the optimal  $\mu$.
    \item We set up the critical coupling regime which is indicated by the nearly maximal depth of the dip, at which the LI-curve width is also maximal.
    \item We adjust the phase so that the LI-curve acquires the correct shape -- the first angle (counting in the frequency scanning direction) should be sharp, and the second -- with a rounding in the scanning direction.
    \item If we do not know the $\mu$ and have not yet selected a mode and/or cannot change the laser position, then we can just look for the resonance with the correct shape (see above).
\end{enumerate}
Any particular experimental realization of the laser requires an adjustment of the optimization algorithm in accordance with the theoretical model described above.

\subsection{Model limitations and nonlinear effects}

There are some limitations on the linewidth reduction with the self-injection locking that are not included into our model.For example, mechanical noise may limit the stability of the self-injection locking. The influence of the mechanical noise on the frequency noise of the self-injection locked laser was partially discussed in \cite{Kondratiev:17}. It was shown that its influence grows with increase of $\tau_s$. The microresonator eigenfrequency noise \cite{Matsko:07} (predominantly thermorefractive noise \cite{KONDRATIEV20182265,Huang2019}) and locking phase variations (due to mechanical as well as thermal instability of the device \cite{photonics5040043} or thermorefractive noises of the lenses and waveguides) will influence the performance of the real device. Thermodynamic noises are not subject to parameter optimization under study and provide a fundamental limit.  

The power accumulation inside the microresonator mode also can result in the stability limitation. The high intracavity intensity can lead to unwanted nonlinear generation effects (e.g. four wave mixing or stimulated Raman scattering) and to the transfer of the laser relative intensity noise (RIN) to the frequency noise. Interestingly, hyper-parametric oscillation (or even soliton generation) \cite{Pavlov_18np,Raja2019} and  Raman lasing \cite{PhysRevLett.105.143903} were observed in the self-injection locking regime. However, their effect on laser characteristics was not thoroughly studied. We presume these effects to be unwanted for a laser frequency stabilization in the self-injection locking regime.
Evidently, their influence could be reduced by decreasing the power inside the microresonator mode. 

The threshold of the parametric-instability related processes can be estimated from the normalized pump expression \cite{herr2014temporal,KondratievBW2019}
\begin{equation}
\label{HPO}
f=\sqrt{\frac{6\chi_3Q_0\eta(1-\eta)^2 P_{\rm input}}{\kappa_0 n^4\epsilon_0 V_0}}\sqrt{\frac{nS_0}{n_cS}}>1,
\end{equation}
where $P_{\rm input}$ is the pump power, $\chi_3$ is microresonator third order nonlinearity, $Q_0=\omega/\kappa_0$ is its internal quality factor, $V_0$ is the mode volume, $n$ and $n_c$ are refraction indices of microresonator and coupler.
The hyper-parametric oscillation and  Raman lasing have nearly identical thresholds in the WGM resonators \cite{PhysRevLett.73.2440,PhysRevA.71.033804,PhysRevLett.105.143903}. In what follows we considered a few approaches reducing the power inside a microresonator mode below the threshold of the hyperparametric oscillations \eqref{HPO}.

The easiest approach to reduce the power circulating in the microresonator mode is to under- or over-couple the microresonator, changing $\eta$. Another solution is to refocus the laser beam in the coupling region (see Fig. \ref{fig:scheme}), changing $\Theta=S_{\rm LC}/S$, the ratio of the laser aperture area $S_{\rm LC}$ to the final beam area $S$. 
The refocusing changes the power of the back-reflected beam by a factor of $\sqrt{\Theta}$ (see Fig. \ref{fig:scheme}), which was hidden in $\tilde\kappa_{do}$ for the simplicity of notation.
In both approaches the stabilization coefficient ($K$) is reduced, thus there exists a trade-off for the conditions of keeping high $K$ and reducing the pump power. 

To consider this problem we introduce the dimensionless ratio:
\begin{equation}
\label{MegaRatio}
\Sigma = \frac{\left.\frac{d K}{d f}\right|_\eta}{\left.\frac{d K}{df}\right|_\Theta} = \frac{\frac{\partial K}{\partial \eta}/\frac{\partial f}{\partial \eta}}{\frac{\partial K}{\partial \Theta}/\frac{\partial f}{\partial \Theta}},
\end{equation}
where $\Theta$ is the effective coupling region area [see Fig. \ref{fig:scheme}]. The numerator of the ratio describes the relative change (speed) of $K$ with respect to $f$ due to $\eta$ tuning (the parametric function derivative $dK/df$). The denominator of the ratio describes the relative change of $K$ with respect of $f$ but with $\Theta$ tuning. Thus, the ratio \eqref{MegaRatio} itself has the meaning of the effectiveness of the tuning $\eta$ over $\Theta$ in terms of keeping high $K$ and reducing $f$. 
 
This ratio is invariant with respect of the redefinition of the stabilization parameter $K\rightarrow F(K)$, where $F$ is an arbitrary function. Thus, a consideration of the the stabilization coefficient $K$ or the linewidth reduction coefficient $K^{-2}$ gives the same optimization result. The same conclusion is true for $f$.

\begin{figure}[ht]
\centering
\includegraphics[width=1.\linewidth]{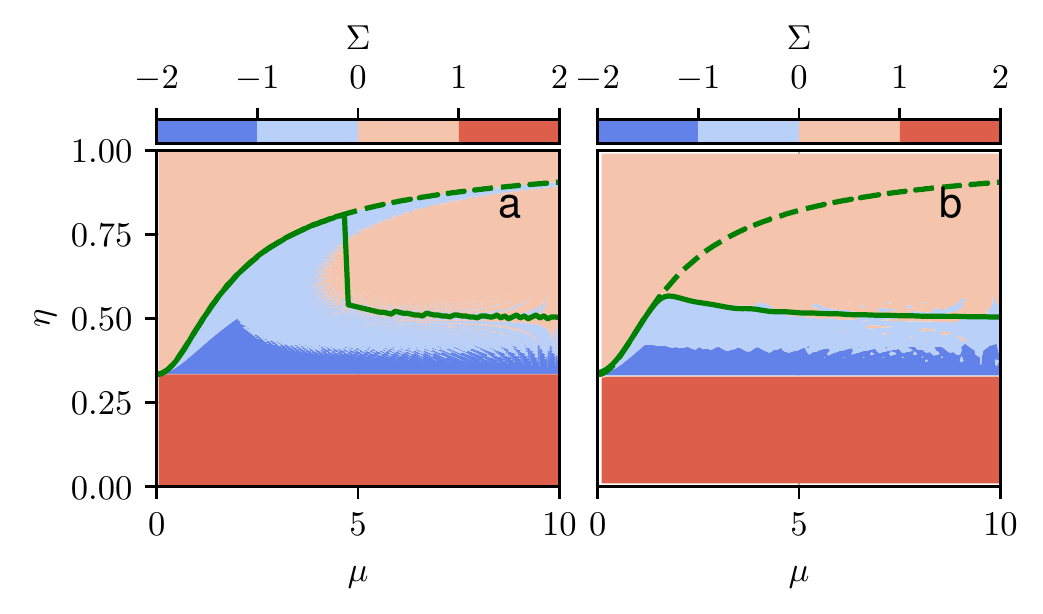}
\caption{a)Parameter $\Sigma$ for the optimal $\zeta_{0}, \ \eta_{0}$ and fixed phase $\psi = 0$; b) Parameter $\Sigma$ for the optimal $\psi_{\rm opt}$,  $\zeta_{\rm opt}, \ \eta_{\rm opt}$. The solid green line stands for the optimal $\eta(\mu)$ trace and dashed green line represents \eqref{eta0}.}
\label{fig:crocodile}
\end{figure}

Figure \ref{fig:crocodile} shows the dependence \eqref{MegaRatio} of the parameter $\Sigma$ on the pump and backward wave coupling values for both zero and the optimal locking phase together with the optimal $\eta(\mu)$ trace. It can be seen that the pump coupling tuning is preferable as the optimal curves are inside $|\Sigma|<1$ region for the both cases.

\section{Conclusion}

We have performed five-parametric optimization of a laser self-injection-locked to high-Q WGM resonator. It was found that the optimal self-injection locking parameters ($\psi , \ \zeta , \ \eta$)  depend only on one parameter of the resonator, $\beta$ (backscattering rate normalized to the microresonator loaded linewidth), and on the distance between the laser and the microresonator (parameter $\kappa_0 \tau_s$). The two optimal parameter regions were found to be separated by $\beta_{\rm cr}\in[0.68;1]$ depending on the distance parameter \eqref{betacr}. 

The optimal combinations of the self-injection locking parameters were derived for different experimentally feasible regimes: all optimal parameters ($\psi_{\rm opt} , \ \zeta_{\rm opt} , \ \eta_{\rm opt}$); long-arm regime $\kappa_0 \tau_s > 0.1$; critical coupling $\eta = 0.5$ regime; fixed phase $\psi=0$, which diverges into two branches corresponding to the regimes of overcoupling and critical coupling.

Both the zero detuning and zero phase are the optimal values for the low backscattering regime. We also found a global maximum of the stabilization coefficient \eqref{global} for the other parameters ($\eta=1/2$, $\beta\approx0.58$).  The optimal parameter values and the stabilization coefficient increase with the distance parameter  $\kappa_0 \tau_s$. The global maximum always stays in the low-$\beta$ region (with $\psi=0$ and $\zeta=0$) as the $\beta_{\rm cr}$ also grows with $\kappa_0\tau_s$. The case of the high-backscattering regime was also discussed.

We have found that for the experimental parameters \cite{Liang2015} at least an order of magnitude improvement of the linewidth reduction is possible if the parameters of the setup are tuned to the optimal point $\eta_{\rm opt}, \ \psi_{\rm opt}$ and $\zeta_{\rm opt}$. The linewidth can be reduced even further, by nearly two orders of magnitude, if the resonator's mode is optimally selected. The recommendations on the experimental realization of the optimal self-injection locking regime were listed. We also have proposed and discussed methods for suppression of the influence of the unwanted nonlinear effects. Our analysis have shown that overcoupling of the resonator mode looks more promising than the geometrical mode mismatch.

\begin{acknowledgments}
The work was supported by the Russian Science Foundation (project 19-72-00173). The reported here research performed by A.M was carried out at the Jet Propulsion Laboratory, California Institute of Technology, under a contract with the National Aeronautics and Space Administration.
\end{acknowledgments}

\appendix

\section{Zero phase optimization}
\label{App1}
We look for the effective detuning $\zeta_0$ that maximizes the stabilization coefficient $K$. Let us introduce $b_{\beta}=\beta^2+1$ for simplicity of notations. The equation 
\begin{equation}
    \frac{\partial K(\zeta,0,\eta,\beta)}{\partial\zeta}=0
    \end{equation}
can be presented as a multiplied by $\zeta$ bi-cubic equation with respect of $\zeta$
\begin{align}
\label{psi0bicubic}
\zeta(3\zeta^6-3(b_{\beta}-2)\zeta^4-(3b_{\beta}^2+8b_{\beta}-8)\zeta^2+\\ \nonumber +3b_{\beta}^2(b_{\beta}-2))=0.
\end{align}
The only free parameter is $\beta$.
The determinant of the cubic expression 
\begin{align}
S =(b_{\beta}-1)(9b_{\beta}^4-6b_{\beta}^3+11b_{\beta}^2-10b_{\beta}+5)
\end{align}
is greater than zero as $\beta>0$ ($b_{\beta}>1$), so the bi-cubic part of \eqref{psi0bicubic} has three root pairs. The first root pair is imaginary for $\beta>0$ and, hence, can be omitted. The second root pair corresponds to the minimum of the stabilization coefficient. Finally, the third root pair corresponds to the maximum of the stabilization coefficient and becomes imaginary for $b_{\beta} < 1$. We are interested in this root. We expand it into series in the vicinity of zero and at infinity. These two approximations intersect at $\beta=1.48$. Combining the solutions we get the expression \eqref{zeta0}.

The approximation for the first part of the $\eta(\mu)$ curve can be found by optimizing the $\beta < 1$ part of the expression \eqref{psi=0} for $\eta$.
By substituting $\beta=\mu(1-\eta)$ into expression \eqref{psi=0} and differentiating it with respect of $\eta$ one derives the following characteristic equation
\begin{align}
\label{eta0eq}
\mu(1-\eta)(\mu^2\eta^3-\mu^2\eta^2-(\mu^2+3)\eta+\mu^2+1)=0.
\end{align}
This equation can be solved for $\mu$:
\begin{align}
\label{muopt0}
\mu=\sqrt{\frac{3\eta-1}{(\eta+1)(1-\eta)^2}}.
\end{align}
It is useful to solve the equation \eqref{eta0eq} for $\eta$. The determinant of the cubic equation \eqref{eta0eq} is positive, so we have three roots, but only one is inside $\eta\in[0;1]$. This root can be expanded into series in the vicinity of $\mu=0$ and $\mu=\infty$ to get the asymptotic expression \eqref{eta0}.

\section{Optimal phase}
\label{App2}
Equation \eqref{stabilization} can be optimized for $\psi$ as follows. Taking the derivative with respect of $\psi$ we derive 
\begin{align}
0=\frac{\partial K}{\partial \psi}&= -a\sin\bar\psi+b\cos\bar\psi, 
\end{align}
where coefficients $a$ and $b$ were introduced in \eqref{stabilization}. Denoting $\sin\alpha=b/(a^2+b^2)^{1/2}$ and using \eqref{masretphase}, we obtain 
\begin{align}
0&=\sin(\psi+\frac{\kappa_m\tau_s}{2}\zeta-\alpha),
\end{align}
solution of which is given by \eqref{psi_0}. This expression can be substituted into Eq.~\eqref{stabilization} to get the phase-optimized stabilization coefficient \eqref{stabilizationpsi}. 

Introducing $W=\kappa_m\tau_s/2=\kappa_0\tau_s/[2(1-\eta)]$ and taking the derivative  with respect of $\zeta$ we get
\begin{align}
\label{zetaopteqass}
0=\zeta(&W^2\zeta^6-(3(W(Wb_{\beta}-2W-2)-2))\zeta^4+\nonumber\\
&+(W(W(3b_{\beta}^2-8b_{\beta}+8)+4b_{\beta}+8)-4b_{\beta}+16)\zeta^2\nonumber\\
&-W(Wb_{\beta}^2(b_{\beta}-2)+2b_{\beta}(5b_{\beta}-8))-2b_{\beta}^2-8b_{\beta}+16).
\end{align}
We note that $\zeta=0$ is an extremum for all the values of $W$. Using $W=0$ we derive a bi-quadratic equation for $\zeta$, which can be easily solved. Equation \eqref{zetaopteqass} can also be solved analytically for arbitrary $W$ as it is a bi-cubic equation. Its discriminant switches sign from positive to negative at $b_{\beta}$, which is decreasing with $W$ increase. Only one of the roots is positive when a specific value of $\beta_{\rm cr}$ is reached. Expanding this root into series with respect of $W$ we derive \eqref{zetaopt} with the following round-trip time correction:
\begin{align}
\label{dzetaopt}
\delta_{\tau_s}\zeta_{\rm opt}=
    \beta^2\frac{\sqrt{\beta^2+3}(\beta^2-1)-\beta(\beta^2+1)}{\sqrt{\beta^2+3}(\beta^4+6\beta^2-3)}\frac{\kappa_0\tau_s}{2(1-\eta)}.
\end{align}
The best fit is provided only with 4-th order series with respect of $\beta$. 

\begin{figure}[ht]
\centering
\includegraphics[width=0.9\linewidth]{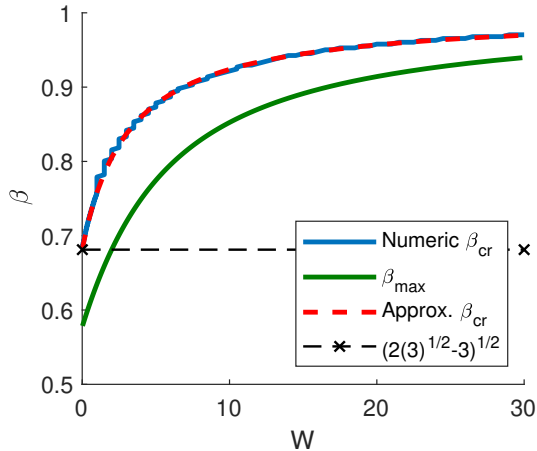}
\caption{The comparison of the exact numerical and the approximated border of the zero detuning optimal parameter selection. The solid blue line is the numerical solution, the red dashed line is the approximation \eqref{betacr} and the solid green line stands for the global maximum $\beta_{max}$ \eqref{global}. The threshold \eqref{zetaopt} $\sqrt{2\sqrt{3}-3}$ is shown by the black dashed line.}
\label{fig:betaregion}
\end{figure}

The exact solution of Eq.~\eqref{zetaopteqass} shows that the threshold value of the backward wave coupling $\beta_{\rm cr}$ (see \eqref{zetaopt}) at which the optimal $\zeta$ becomes nonzero changes with $W$.
The exact analytical solutions in the special cases show that $\beta_{\rm cr}=\sqrt{2\sqrt{3}-3}$ for $W=0$ and $\beta_{\rm cr}\rightarrow1$ for $W\rightarrow\infty$. 
The approximation of $\beta_{\rm cr}(W)$ can be constructed as a rational function \eqref{betacr} having the above limits. This approximation was tested numerically and exhibited a very good correspondence with the exact solution of \eqref{zetaopteqass}. The figure \ref{fig:betaregion} shows the approximation \eqref{betacr} (red-dashed line) together with the numerical estimation (solid blue line). The dependence of optimal backscattering $\beta_{\rm max}$ \eqref{global} is also presented to show that the global maximum is always bellow $\beta_{\rm cr}$, i.e. in the zero detuning region. 

We also derived an asymptotic solution for the large $\beta$
\begin{align}
\label{zetaoptbinf}
\zeta_{\rm opt}=\beta-\frac{W(W+2)}{2(W^2+4W+2)\beta}+O(\beta^{-2}).
\end{align}

To get the global maximum we substitute $\zeta=0$ \eqref{zetaopt} into \eqref{stabilizationpsi} and obtain \eqref{Kzopt}. 
By taking derivatives with respect of $\beta$ and $\eta$ and solving corresponding equations we obtain
\begin{align}
\label{bopteta}
\beta_{\rm o1}=&\sqrt{\frac{6(\eta-1)+\sqrt{32(1-\eta)^2+(2(1-\eta)+\kappa_0\tau_s)^2}}{\kappa_0\tau_s}},\\
\label{eoptbeta}
\eta_{\rm o1}=&\frac{\kappa_0\tau_s}{8}(\beta^2+1)+\frac{1}{2}.
\end{align}
Inserting \eqref{eoptbeta} into the derivative of \eqref{Kzopt} with respect to $\beta$ we get bi-quadratic equation for $\beta_{\rm max}$. Inserting the result into \eqref{eoptbeta} and into \eqref{Kzopt} we derive \eqref{global}.

\label{App3}
To obtain $\eta_{\rm opt}$ we substitute  $\beta=\mu(1-\eta)$ into \eqref{stabilizationpsi} and take a derivative with respect of $\eta$
\begin{align}
\label{etaopt0eq}
0=\mu(1-\eta)(&2(1-\eta)^2(\kappa_0\tau_s+2\eta+2)\mu^2-4(3\eta-1)+\nonumber\\
&+\kappa_0\tau_s(1-\eta)^3\mu^4-\kappa_0\tau_s\frac{2\eta-1}{1-\eta}).
\end{align}
Though this is a complete 4$^{th}$-order equation with respect of $\eta$, it is a bi-quadratic equation with respect to $\mu$ and, hence, can be solved analytically. The only real root is
\begin{align}
\label{muopt}
\mu_{\rm opt}=&\frac{1}{\sqrt{(1-\eta)^3W}}\Big(W(2\eta-1)-\eta-1+\nonumber\\
                &+\sqrt{W^2\eta^2-2\eta(5\eta-3)W+(\eta+1)^2}\Big)^{1/2}.
\end{align}
The rough estimation of the correction for $\eta_{0}$ can be derived for small $\mu$. Neglecting $\mu^2$ in \eqref{etaopt0eq} we obtain
\begin{align}
12\eta^2-(2\kappa_0\tau_s+16)\eta+\kappa_0\tau_s+4=0.
\end{align}
The bigger root exceeds unity and thus is outside of the $\eta$ validity range. Expanding the smaller root into series we derive
\begin{align}
\eta_{\beta=0}\approx\frac{1}{3}+\frac{\kappa_0\tau_s}{24}-\frac{\kappa_0^2\tau_s^2}{128}.
\end{align}
Combining this expression with \eqref{eta0} and assuming small $\beta$ (or $\mu$) we get the low-scattering limit of \eqref{etaopt}.

\bibliography{sample}

\end{document}

%% file: preamble.tex
\usepackage{amsthm}
\usepackage{mathtools}
\usepackage{physics}
\usepackage{xcolor}
\usepackage{graphicx}
\usepackage[left=23mm,right=13mm,top=35mm,columnsep=15pt]{geometry} 
\usepackage{adjustbox}
\usepackage{placeins}
\usepackage[T1]{fontenc}
\usepackage{lipsum}
\usepackage{csquotes}